\documentclass[aps,prd,onecolumn,groupedaddress,showpacs,nofootinbib,amssymb]{revtex4}
\usepackage[dvipdfmx]{graphicx}
\usepackage[dvipdfmx]{color}
\usepackage{bm}
\usepackage{amsmath}
\usepackage{amssymb}
\usepackage{amsfonts}
\usepackage{slashbox}
\usepackage{comment}

\newcommand{\be}{\begin{equation}}
\newcommand{\ee}{\end{equation}}
\newcommand{\bea}{\begin{eqnarray}}
\newcommand{\eea}{\end{eqnarray}}
\newcommand{\beaa}{\begin{eqnarray*}}
\newcommand{\eeaa}{\end{eqnarray*}}

\newcommand{\nn}{\nonumber \\}
\newcommand{\e}{\mathrm{e}}

\allowdisplaybreaks[4]

\begin{document}

\title{Stability analysis for new theories of massive spin-two particle and black hole entropy of new bigravity}

\author{Yuichi Ohara$^1$, Satoshi Akagi$^1$ and Shin'ichi Nojiri$^{1,2}$}

\affiliation{
$^1$ Department of Physics, Nagoya University, Nagoya
464-8602, Japan \\
$^2$ Kobayashi-Maskawa Institute for the Origin of Particles and
the Universe, Nagoya University, Nagoya 464-8602, Japan 
}

\begin{abstract}
{
In Phys.\ Rev.\ D 90 043006 (2014), we proposed a new ghost-free massive spin-two model in flat spacetime. 
Furthermore, as some extension, we couple the new model with a non-dynamical curved background in Phys.\ Rev.\ D 90 123013 (2014) 
and constructed new interaction terms without appearance of an extra mode. The characteristic property of the new model is the 
existence of nonlinear potential terms which give the nontrivial vacua. The presence of the nontrivial vacua, however, does not mean 
that the particle can be defined around all vacua. Therefore, in this paper, we discuss the condition for the new model to have stable vacua 
in flat spacetime and curved spacetime. Then, we couple this spin-two theory with a dynamical background and obtain the solutions.
Moreover, we investigate the effect of this new spin-two model to the Einstein gravity by calculating the black hole entropy 
since the gravity coupled with massive spin-two theory admits a black hole solution in addition to (anti-) de Sitter space solution.
}
\end{abstract}


\maketitle

\section{Introduction \label{Sec1}}

The consistent free massive spin-two theory was first established by Fierz and 
Pauli \cite{Fierz:1939ix}. 
The mass term for spin-two particles generally leads to a ghost mode, but they preserve 
the consistency of the theory by tuning coefficients of the mass term. 
Since the Fierz-Pauli theory does not have any gauge symmetry, it seems that arbitrary 
interactions can be added to the theory. 
Contrary to this naive expectation, Boulware and Deser \cite{Boulware:1974sr} showed 
that non-linear terms generally lead to another ghost called the Boulware-Deser (BD) ghost. 
There was another problem, that is, the appearance of the van Dam-Veltman-Zakharov 
(vDVZ) discontinuity \cite{vanDam:1970vg} in the massless limit, $m\to 0$ 
although the discontinuity can be screened 
by the Vainstein mechanism \cite{Vainshtein:1972sx} 
(see, for example, Ref.~\cite{Luty:2003vm}).

After these indications, the studies of 
massive spin-two fields had not progressed until 2003.
In 2003, Arkani-Hamed, Georgi, and Schwartz \cite{ArkaniHamed:2002sp}, however, 
revealed a cutoff scale of the theory by introducing the Stuckelberg field. 
They considered a limit which focuses on the cutoff and they have shown that the special 
choice of the coefficients in the potential terms makes the cutoff scale larger. 
As the potential-tuned theory consists of infinite terms, it was unclear whether the 
theory contains the BD ghost or not. After that, de Rham, Gabadadze and Tolley \cite{deRham:2010ik,deRham:2010kj} 
succeeded in the resummation of the potential terms and Hassan and Rosen \cite{Hassan:2011hr} proved 
that the theory with the resummed potential terms does not contain any ghost. 
This theory is called the dRGT massive gravity. The most important point in this 
theory is special forms of the fully non-linear potential terms eliminating the extra mode.
Although the massive gravity models have non-dynamical background metric, the models 
have been extended to the models with dynamical metric
\cite{Hassan:2011zd,Hassan:2011vm,Hassan:2011tf}, 
which are called as bigravity models. 

Hinterbichler \cite{Hinterbichler:2013eza} (see also \cite{Folkerts:2011}) pointed out the possibility of new 
derivative interaction terms in the dRGT massive gravity. 
It was shown that new derivative interactions can be added to the Fierz-Pauli theory by 
the taking specific linear combination of interactions and conjectured fully 
non-linear counterparts of these interaction terms in dRGT massive gravity. 
In this context, it was also shown that the leading term of the dRGT potential term does 
not generate the ghost to the Fierz-Pauli theory. 
Thus, we constructed a new massive spin-two model in a flat spacetime by adding the leading terms to 
Fierz-Pauli free theory in \cite{Ohara:2014vua}. Furthermore, we extend the theory to the rigid 
curved background and show that the theory is ghost-free on Einstein manifold \cite{Akagi:2014vua}.

In this paper, we investigate the stability of the potential extrema of the new model in flat spacetime and curved spacetime.
Furthermore, we consider the model where the field of massive spin-two particle couples 
with gravity by assuming, for simplicity, that 
the spin-two field is proportional to the background metric. The other kind of solutions have been found \cite{Volkov2012tf, Maeda2013tf, 
Volkov:2011an} in the context of 
the Hassan-Rosen bigravity model\cite{Hassan:2011zd,Hassan:2011vm,Hassan:2011tf}.
A reason why we consider this model is an application to the cosmology and black hole (BH) 
physics.  We often consider the models of scalar fields to explain the expanding universe 
not to violate the isotropy while the condensation of the vector field violate 
the isotropy in general except the case that the model has a non-abelian gauge 
symmetry.\footnote{
Non-abelian gauge always contains $SU(2)$ or $SO(3)$ as a subgroup. 
The condensation of the vector field breaks both of the isotropy or rotational 
invariance and the gauge symmetry. 
Because the rotational symmetry is $SO(3)$, even if the vector field condensate, 
there remain the diagonal symmetry in the product of the rotational symmetry $SO(3)$ 
times the guage symmetry $SO(3)$ and we can regard the diagonal symmetry as a 
new rotational symmetry. 
} 
The field of the massive spin-two particle is given by rank 2 symmetric tensor. 
We should note that the condensation of the trace part of the rank 2  
symmetric tensor (or $(t,t)$ component, or the trace of the spacial part) does not violate 
the isotropy and therefore we can use the rank 2 symmetric tensor in order to explain 
the expansion of the universe. Such a cosmology has been studied in the massive gravity models \cite{deRham:2010tw} 
by considering the decoupling limit where the models reduce to scalar-tensor theories. 
After that there follow several activities in the massive gravity models 
\cite{Kluson:2012zz,Kluson:2012wf,Hassan:2011ea,D'Amico:2011jj} 
and in the bimetric gravity models 
\cite{Damour:2002wu,Volkov:2011an,vonStrauss:2011mq,
Berg:2012kn,Nojiri:2012zu,Nojiri:2012re,Bamba:2013fha,AKMS-TSK}. 

{
As for black hole physics, the effect of massive spin-two particles to the black hole 
entropy has been already calculated in the Hassan-Rosen bigravity model \cite{Katsuragawa:2013bma,Katsuragawa:2013lfa}.
Since the gravity coupled with massive spin-two model presented in this paper is essentially different from the Hassan-Rosen bigravity model 
\cite{Hassan:2011zd,Hassan:2011vm,Hassan:2011tf}, it is quite interesting to see how the result change depending on the model.
}

\section{New model of massive spin-two particle}

The Lagrangian of the Fierz-Pauli theory is given by \cite{Fierz:1939ix} 
\be
\label{FPLag}
\mathcal{L}_{\mathrm{FP}} = -\frac{1}{2}\partial_\lambda h_{\mu\nu}\partial^\lambda 
h^{\mu\nu}+\partial_\mu h_{\nu\lambda}\partial^\nu h^{\mu\lambda}-\partial_\mu 
h^{\mu\nu}\partial_\nu h+\frac{1}{2}\partial_\lambda h\partial^\lambda h 
 -\frac{1}{2}m^2 \left( h_{\mu \nu}h^{\mu \nu}-h^2 \right) \, .
\ee
The relative sign of the mass term is tuned to eliminate a ghost. 
Hinterbichler pointed out that new interaction terms can be added to this model
without any ghost by taking the specific linear combination \cite{Hinterbichler:2013eza, Folkerts:2011}. 
In four dimensions, there are two kinds of non-derivative interactions :
\begin{align}
\mathcal{L}_3 
\sim & \eta^{\mu_{1} \nu_{1} \mu_{2} \nu_{2} \mu_{3} \nu_{3}} h_{\mu_{1}\nu_{1}} h_{\mu_{2} 
\nu_{2}} h_{\mu_{3} \nu_{3}} \, ,\label{nhhh} \\
\mathcal{L}_4
\sim & \eta^{\mu_{1} \nu_{1} \mu_{2} \nu_{2} \mu_{3} \nu_{3} \mu_{4} \nu_{4}} 
h_{\mu_{1} \nu_{1}} h_{\mu_{2} \nu_{2}} h_{\mu_{3} \nu_{3}} h_{\mu_{4} \nu_{4}} \, .
\label{nhhhh}
\end{align}
Here $\eta^{\mu_{1} \nu_{1} \cdots \mu_{n} \nu_{n}}$ is given by the product of $n$ 
$\eta_{\mu\nu}$ and anti-symmetrizing the indexes $\nu_1$, $\nu_2$, $\cdots$, and 
$\nu_n$, for examples, 
\begin{align}
\label{h3c}
\eta^{\mu_{1} \nu_{1} \mu_{2} \nu_{2}} \equiv &  
\eta^{\mu_{1} \nu_{1}} \eta^{\mu_{2} \nu_{2}} - \eta^{\mu_{1} \nu_{2}} \eta^{\mu_{2} 
\nu_{1}}\, , \nn
\eta^{\mu_{1} \nu_{1} \mu_{2} \nu_{2} \mu_{3} \nu_{3}} \equiv & 
\eta^{\mu_{1} \nu_{1}}\eta^{\mu_{2} \nu_{2}} \eta^{\mu_{3} \nu_{3}} - \eta^{\mu_{1} 
\nu_{1}}\eta^{\mu_{2} \nu_{3}} \eta^{\mu_{3} \nu_{2}}
+ \eta^{\mu_{1} \nu_{2}}\eta^{\mu_{2} \nu_{3}} \eta^{\mu_{3} \nu_{1}} \nn
& - \eta^{\mu_{1} \nu_{2}}\eta^{\mu_{2} \nu_{1}} \eta^{\mu_{3} \nu_{3}}
+ \eta^{\mu_{1} \nu_{3}}\eta^{\mu_{2} \nu_{1}} \eta^{\mu_{3} \nu_{2}} - \eta^{\mu_{1} 
\nu_{3}}\eta^{\mu_{2} \nu_{2}} \eta^{\mu_{3} \nu_{1}} 
\, .
\end{align}
In \cite{Ohara:2014vua}, we proposed the new model of massive spin-two particles by adding 
the two terms (\ref{nhhh}) and (\ref{nhhhh}) to the Fierz-Pauli Lagrangian.
\begin{align}
\label{hh10}
\mathcal{L}_{h0} 
= & -\frac{1}{2} \eta^{\mu_{1} \nu_{1} \mu_{2} \nu_{2} \mu_{3} \nu_{3}} 
\left( \partial_{\mu_{1}} \partial_{\nu_{1}} h_{\mu_{2} \nu_{2}}\right) h_{\mu_{3} \nu_{3}}
 + \frac{m^2}{2} \eta^{\mu_{1} \nu_{1} \mu_{2} \nu_{2}} h_{\mu_{1} \nu_{1}} h_{\mu_{2} 
\nu_{2}} \nn
& - \frac{\mu}{3!} \eta^{\mu_{1} \nu_{1} \mu_{2} \nu_{2} \mu_{3} \nu_{3}} 
h_{\mu_{1}\nu_{1}} h_{\mu_{2} \nu_{2}} h_{\mu_{3} \nu_{3}}
 - \frac{\lambda}{4!} \eta^{\mu_{1} \nu_{1} \mu_{2} \nu_{2} \mu_{3} \nu_{3} \mu_{4} \nu_{4}} 
h_{\mu_{1} \nu_{1}} h_{\mu_{2} \nu_{2}} h_{\mu_{3} \nu_{3}} h_{\mu_{4} \nu_{4}}
\, \nn
= & -\frac{1}{2} \left( h \Box h - h^{\mu\nu} \Box h_{\mu\nu} 
 - h \partial^\mu \partial^\nu h_{\mu\nu} - h_{\mu\nu} \partial^\mu \partial^\nu h 
+ 2 h_\nu^{\ \rho} \partial^\mu \partial^\nu h_{\mu\rho} \right) \nn
& + \frac{m^2}{2} \left( h^2 - h_{\mu\nu} h^{\mu\nu} \right) 
 - \frac{\mu}{3!} \left( h^3 - 3 h h_{\mu\nu} h^{\mu\nu} + 2 h_\mu^{\ \nu} h_\nu^{\ \rho} 
h_\rho^{\ \mu} \right)  \nn
& - \frac{\lambda}{4!} \left( h^4 - 6 h^2 h_{\mu\nu} h^{\mu\nu} + 8 h h_\mu^{\ \nu} 
h_\nu^{\ \rho} h_\rho^{\ \mu} 
 - 6 h_\mu^{\ \nu} h_\nu^{\ \rho} h_\rho^{\ \sigma} h_\sigma^{\ \mu} + 3 \left( 
h_{\mu\nu} h^{\mu\nu} \right)^2 \right) \, .
\end{align}
Here $m$ and $\mu$ are parameters with the dimension of mass and $\lambda$ is 
a dimensionless parameters. We assume that $\mu$ always takes a positive value but 
cannot decide the sign of $\lambda$, because it is non trivial 
to learn which sign for $\lambda$ stabilizes this system. 

Although the model (\ref{hh10}) is power counting renormalizable, the model is not 
renormalizable because the propagator behaves as $\mathcal{O}\left( p^2 \right)$ for 
large momentum $p$ instead of the naive expectation $\mathcal{O}\left( p^{-2} \right)$. 
In fact, the propagator has the following form: 
\begin{align}
\label{hh1}
D^m_{\alpha\beta,\rho\sigma} =& - \frac{1}{2\left( p^2 + m^2 \right)}
\left\{  P^m_{\alpha\rho} P^m_{\beta\sigma} + P^m_{\alpha\sigma} P^m_{\beta\rho}  
 - \frac{2}{3} P^m_{\alpha\beta} P^m_{\rho\sigma} \right\} \, , \\
\label{hh2}
P^m_{\mu\nu} \equiv& \eta_{\mu\nu} + \frac{p_\mu p_\nu}{m^2} \, .
\end{align}
Then when $p^2$ is large, the propagator behaves as  
$D^m_{\alpha\beta,\rho\sigma} \sim \mathcal{O}\left( p^2 \right)$ 
due to the projection operator $P^m_{\mu\nu}$, which makes the behavior for large
$p^2$ worse and therefore the model should not be renormalizable. 

Since this theory has no symmetry and is already non-renormalizable, it seems that there 
is no reason why we only consider the potential terms up to the quartic order. 
However, introducing higher order potential terms break the consistency as quantum field 
theory in four dimensions. 
The potential terms described above does not generate any ghost due to the 
anti-symmetric property. 
Therefore, in four dimensions, we cannot construct similar ghost-free potential terms. 
Needless to say, we can add higher order terms in five or higher dimensions.

\section{Classical solution in new theory of massive spin-two field} 

Because the potential of the new theory of massive spin-two field has a structure like the 
potential of the Higgs field, it could be interesting to investigate the classical solutions, 
which may correspond to the extrema of the potential. 
The non-vanishing value of the potential for the classical solution may give an energy of 
the vacuum.  

By the variations of $h_{\mu \nu}$, we obtain the equations of motion for $h_{\mu \nu}$, 
\begin{align}
\frac{\delta S}{\delta h^{\mu\nu}}=&
\Box h_{\mu\nu}-\partial_\lambda\partial_\mu h^\lambda_{\ \nu}
 -\partial_\lambda\partial_\nu 
h^\lambda_{\ \mu}+g_{\mu\nu}\partial_\lambda\partial_\sigma h^{\lambda\sigma}
+\partial_\mu\partial_\nu h - g_{\mu\nu}\Box h \nonumber \\
&-m^2(h_{\mu\nu}-g_{\mu\nu}h)-\frac{\mu}{3!} (3 g_{\mu \nu} h^2 - 3 g_{\mu \nu} 
h_{\rho \sigma}h^{\rho \sigma} - 6hh_{\mu \nu}+3{h_{\nu}}^{\rho}
h_{\rho \mu}+3{h_{\mu}}^{\rho}h_{\rho \nu}) \nonumber \\ 
&-\frac{1}{4!}( 4 g_{\mu \nu}h^3 -12 g_{\mu \nu} h h_{\rho \sigma}h^{\rho \sigma} -12 h^2 
h_{\mu \nu}+ 8 g_{\mu \nu} {h^{\rho}}_{\sigma} {h^{\sigma}}_{\kappa} {h^{\kappa}}_{\rho} 
+12h {h_{\nu}}^{\rho}h_{\rho \mu}+12 h {h_{\mu}}^{\rho} h_{\rho \nu} \nonumber \\
&-12{h_{\mu}}^{\rho}{h_{\rho}}^{\sigma}h_{\sigma \nu}-12{h_{\nu}}^{\rho}
{h_{\rho}}^{\sigma}h_{\sigma \mu}+12 h_{\mu \nu} (h^{\rho \sigma}h_{\rho \sigma}) )
=0 \, .
\label{eom}
\end{align}
We assume the solution of equations (\ref{eom}) is given by 
\be
\label{hhhh2}
h_{\mu\nu} = C \eta_{\mu\nu}\, .
\ee
Here $C$ is a constant. Substituting (\ref{hhhh2}) into the equations (\ref{eom}) gives
\be
(3m^2C-3\mu C^2-\lambda C^3) \eta_{\mu \nu}=0 \, .
\label{eom2}
\ee
The solutions for (\ref{eom2}) are given by
\be
\label{sol}
C=0\, , \quad \frac{-3 \mu \pm \sqrt{9 \mu^2 + 12 m^2 \lambda}}{2 \lambda}\, .
\ee
Because the solution should be a real number, the parameters are constrained to be
\[
\left\{
\begin{array}{l}
\lambda \geq -\frac{3\mu^2}{4 m^2} \quad \mbox{for} \quad m^2 > 0\\
\\
\lambda \leq \frac{3 \mu^2}{4 |m^2|} \quad \mbox{for} \quad m^2 <0  \\
\end{array}
\right. \, .
\]
Note that the parameter $m^2$ is not required to be positive definite due to the presence of the potential terms.
By assuming (\ref{hhhh2}), the Lagrangian (\ref{hh10}) is reduced to 
\be
\label{R19}
\mathcal{L}_{h0} = V(C) \equiv - 6m^2 C^2 + 4\mu C^3 + \lambda C^4\, .
\ee
We may regard $V(C)$ as a potential for $C$. 
Then Eq.~(\ref{eom2}) is nothing but the condition $V'(C)=0$. 
We should note that when $\mu=\lambda=0$, which corresponds to the Fierz-Pauli model, 
the potential $V(C)$ is not unbounded below and $C=0$ corresponds to the local 
maximum instead of the local minimum. 
As we know, however, that the massive spin-two field is stable on the local maximum. 
On the other hand, on the local minimum of $C$, the fluctuation of the massive spin-two 
field becomes tachyonic and unstable. 

Such a contradiction to the intuition occurs because $C$ does not correspond to the 
propagating mode and $C$ should be a constant. 
In fact, if we assume (\ref{hhhh2}) and that $C$ could not be a constant, (\ref{eom}) tells 
\be
\label{R23}
0 = \eta^{\mu\nu} \left( 2 \Box C + 3 m^2 C - 3 \mu C^2 - \lambda C^3 \right)
 - 2 \partial^\mu \partial^\nu C\, .
\ee
Then when $\mu\neq\nu$ in (\ref{R23}) gives
\be
\label{R24}
\partial_\mu \partial_\nu C = 0\, ,
\ee
which tells that $C$ is given by a sum of the functions of each of coordinates 
$C = \sum_{\mu} C^{(\mu)} (x^\mu)$. 
Eq.~(\ref{R23}) also gives 
\be
\label{R25}
\eta^{\mu\mu} \partial_\mu^2 C = \eta^{\nu\nu} \partial_\nu^2 C\, .
\ee
In Eq.~(\ref{R25}), the indeces $\mu$ in the l.h.s. and $\nu$ in the r.h.s. are not 
summed up. 
Eq.~(\ref{R25}) tells that $C$ takes the following form, 
$C = \sum_{\mu, \nu} \frac{c}{2} \eta_{\mu \nu} x^{\mu}x^{\nu}+\sum_{\mu} c_\mu x^\mu + C_0$. Here $c$, $c_\mu$'s and $C_0$ are constants.  
By substituting this expression into (\ref{R23}), we find $c=0$ and $c_\mu = 0$, which means $C$ 
should be surely a constant. 
This tells that even if $C$ is on the local maximum of the potential (\ref{R19}), 
$C$ does not roll down. 

If $V(C)$ does not vanish, the potential $V(C)$ could be the vacuum energy and might 
play the role of  the cosmological constant when we couple the model with gravity.  
Then it could be interesting to investigate the signature of the potential and the 
(in)stability of the classical solution corresponding to the extrema of the potential. 

We now assume the parameter $\mu$ is positive because the signature of $\mu$ can be 
always absorbed into the redefinition of $h_{\mu\nu}$.
As  the sign of $\lambda$ and $m^2$ are undetermined while $\mu$ takes a positive value, we consider 
the following cases.
{
\begin{description}
\item[(a)] $\lambda >0$ and $m^2 > 0$ case. \\ \\
Besides the trivial solution $C=0$, there are non-trivial solutions for $C$, which are given by
\be
C_{1}=\frac{-3 \mu + \sqrt{9 \mu^2+12m^2 \lambda}}{2 \lambda}>0\, , \quad 
C_{2}=\frac{-3 \mu - \sqrt{9 \mu^2+12m^2 \lambda}}{2 \lambda}<0 \, .
\label{ooh3}
\ee
We now consider which solution corresponds to the positive (negative) energy solution 
under the assumption $\mu>0$, $\lambda > 0$, and $m^2>0$. 
For this purpose, we have to solve inequalities 
\begin{align}
V(C) =&  -6m^2 C^2 + 4\mu C^3 +\lambda C^4 > 0\, ,  \nn
V(C) =&  -6m^2 C^2 + 4\mu C^3 +\lambda C^4 < 0\, . 
\end{align}
The solutions are given by 
\begin{align}
C<C_{-} \quad \mathrm{or} \quad C_{+} <C \quad & \mbox{for} \quad 
\mbox{Positive energy} \, , \label{plambda1} \\ 
 C_{-} < C < C_{+} \quad & \mbox{for} \quad 
\mbox{Negative energy} \, .
\label{nlambda2}
\end{align}
Here $C_{+}$ and $C_{-}$ are defined by 
\be
\label{ooh4}
C_{+}=\frac{-2 \mu + \sqrt{4 \mu^2 + 6 \lambda m^2}}{\lambda}>0\, , \quad 
C_{-}=\frac{-2 \mu -\sqrt{4 \mu^2 + 6 \lambda m^2}}{\lambda}<0 \, .
\ee
In order that $C_{\pm}$ could be real numbers, $\lambda$ should be larger than 
$-2 \mu^2 / 3m^2$, 
but we assume the positivity of $\lambda$ and $m^2$ here. 
$C_{1}$ and $C_{+}$ are both positive and  $C_{2}$ and $C_{-}$ are both negative. 
Thus, what we should do is to compare $C_{1}$ to $C_{+}$ and $C_{2}$ to $C_{-}$, 
respectively. 
\begin{enumerate}
\item $C_{+}$ and $C_{1}$ \\
We now consider the following quantity,
\begin{align}
C_{+}-C_{1} &= \frac{\mu}{2 \lambda} \left(-1 
+ 4 \sqrt{1 +\frac{3m^2 \lambda }{2 \mu^2}}
 -3 \sqrt{1 +\frac{4m^2 \lambda }{3 \mu^2}} \right) \nonumber \\
&> \frac{\mu}{2 \lambda} \left(-1 
+ \sqrt{1 +\frac{4m^2 \lambda }{3 \mu^2}} \right) > 0 \, .
\end{align}
Thus, we obtain the relation $C_+ > C_1$. Since $C_+$ is positive while $C_-$ is negative, 
we find $C_- < C_1 < C_+$, which means the solution
$C_1$ corresponds to the negative energy.
\item $C_{2}$ and $C_{-}$ \\
Similarly, we investigate the difference between $C_-$ and $C_2$, 
\begin{align}
C_- - C_2 &= \frac{\mu}{2 \lambda}\left(-1
 - 4\sqrt{1+\frac{3m^2 \lambda }{2 \mu^2}}+3\sqrt{1
+\frac{4m^2 \lambda }{3 \mu^2}}\right) \nonumber \\
& < \frac{\mu}{2 \lambda} \left(-1-\sqrt{1+\frac{3m^2 \lambda }{2 \mu^2}} \right) <0\, .
\label{qq2}
\end{align}
This means $C_- < C_2$. 
Since $C_{2}$ has a negative value,  $C_{-} < C_{2} < C_+$ is held. 
\end{enumerate}
Therefore, we see that both of solutions satisfy the condition for the negative energy 
solution.
\item[(b)] $-3\mu^2/4m^2<\lambda < 0$  and $m^2>0$ case. \\ \\
We continue the similar analysis. 
However, $C_{1,2}$ and $C_{\pm}$ are given as follows in this case. 
\begin{align}
&C_1 = \frac{3 \mu - \sqrt{9\mu^2 -12 m^2 |\lambda|}}{2 |\lambda|} >0 \, , \quad 
C_2 = \frac{3 \mu + \sqrt{9\mu^2 -12 m^2 |\lambda|}}{2 |\lambda|} >0 \, ,\nn
&C_- = \frac{2 \mu - \sqrt{4\mu^2 -6 m^2 |\lambda|}}{|\lambda|} >0 \, , \quad 
C_+ = \frac{2 \mu + \sqrt{4\mu^2 -6 m^2 |\lambda|}}{|\lambda|} >0 \, , \label{ooh5} \\
&\phantom{C_+ = \frac{2 \mu + \sqrt{4\mu^2 -6 m^2 |\lambda|}}{|\lambda|}} C_{2}>C_{1}\, , \quad C_{-}< C_{+} \, .\nonumber 
\end{align}
Note that $C_-$ and $C_+$ in (\ref{ooh5}) correspond to $C_+$ and $C_-$ in (\ref{ooh4}), respectively.
Since $\lambda$ is the coefficient of $C^4$, the solutions for the inequality also change.
\begin{align}
C_{-} < C < C_{+} \quad &\mathrm{for} \quad \mbox{Positive energy} \, ,
\label{nlambda3} \\
C<C_{-} \  \mathrm{or} \ C_{+} <C \quad &\mathrm{for} \quad 
\mbox{Negative energy} \, .\label{plambda4}
\end{align}
The condition for $C_{\pm}$ to be real is given by $ -2 \mu^2 / 3m^2< \lambda <0 $. 
As we assume the reality of $C$ in this analysis, $C_{\pm}$ does not exist for the case $-3 \mu^2 / 
4m^2 < \lambda < -2 \mu^2 / 3 m^2$. Therefore, we divide the parameter region $-3m^2/4\mu^2<\lambda < 0$ into 
$-3\mu^2/4m^2<\lambda < -2\mu^2/3m^2$ and $-2\mu^2/3m^2< \lambda < 0$. 
\\ \\
\textbf{(b-1)} $-2\mu^2/3m^2< \lambda < 0$ and $m^2>0$ case \\
Let us compare $C_{1,2}$ with $C_{\pm}$.
\begin{enumerate}
\item $C_{-}$ and $C_{1}$ \\
As in the previous case, we consider the quantity
\begin{align}
C_{-}-{C_1}&= \frac{\mu}{2 |\lambda|}\left(1 - 4\sqrt{1
 -\frac{3m^2 |\lambda| }{2 \mu^2}} + 3\sqrt{1-\frac{4m^2 |\lambda| }{3 \mu^2}}\right) \nonumber \\
&> \frac{\mu}{2 |\lambda|} \left(1-\sqrt{1-\frac{3m^2 |\lambda| }{2 \mu^2}}  \right) >0
\, .\label{qqq}
\end{align}
Thus, we find $C_{1}<C_{-}<C_{+}$ and $C_1$ is turned out to be the negative energy solution.
\item $C_{-}$ and $C_{2}$
\begin{align}
C_{-}-{C_2} &= \frac{\mu}{2 |\lambda|}\left(1 - 4 \sqrt{1-\frac{3m^2 |\lambda| }{2 \mu^2}} -3\sqrt{1
 -\frac{4m^2 |\lambda| }{3 \mu^2}}\right) \nonumber \\
&< \frac{\mu}{2 |\lambda|}\left(1 - 4 \sqrt{1-\frac{3m^2 |\lambda| }{2 \mu^2}} -3 \cdot \frac{1}{3} \right) < 0 \, .
\label{qqq2}
\end{align}
In the second line, we use the fact $1/3 <\sqrt{1-{4m^2 |\lambda|} /{3 \mu^2}} < 1$. From (\ref{qqq2}), we obtain the result
$C_-<C_2$.
\item $C_{+}$ and $C_{2}$
\begin{align}
C_{+}-C_2&= \frac{\mu}{2 |\lambda|}\left(1 +4 \sqrt{1 -\frac{3m^2 |\lambda| }{2 \mu^2}} -3\sqrt{1-\frac{4m^2 |\lambda| }{3 \mu^2}}\right) \nonumber \\
&> \frac{\mu}{2 |\lambda|}\left\{1 +3 \left( \sqrt{1 -\frac{3m^2 |\lambda| }{2 \mu^2}} -\sqrt{1-\frac{4m^2 |\lambda| }{3 \mu^2}}\right)   \right\}
\, .\label{qqq3}
\end{align}
The quantity $ \sqrt{1 -{3m^2 |\lambda| }/{2 \mu^2}} -\sqrt{1-{4m^2 |\lambda| }/{3 \mu^2}}$ is always negative if we assume \textbf{(b-1)}. Thus 
(\ref{qqq3}) is rewritten as follows :
\be
C_{+}-C_2 > \frac{\mu}{2 |\lambda|}\left\{1 -3 \left| \sqrt{1 -\frac{3m^2 |\lambda| }{2 \mu^2}} -\sqrt{1-\frac{4m^2 |\lambda| }{3 \mu^2}}\right|   \right\} \, .
\ee
The maximum value of the second term is given by
\be
\left| \sqrt{1 -\frac{3m^2 |\lambda| }{2 \mu^2}} -\sqrt{1-\frac{4m^2 |\lambda| }{3 \mu^2}}\right| < \frac{1}{3} \, ,
\label{thy}
\ee
in the assumed parameter region. Therefore, $C_+$ is larger than $C_2$.
\end{enumerate}
According to these analysis, $C_2$ and $C_1$ correspond to the positive energy 
solution and the negative energy solution, respectively. \\ \\
\textbf{(b-2)} $ -3\mu^2/4m^2 < \lambda< -2\mu^2/3m^2$ case \\ \\
As mentioned above, $C_{\pm}$ is no longer real in this case. Thus, 
$V(C)$ takes a negative values only,which means that $C_{1}$ and $C_{2}$ produce the negative energy solution.

\item[(c)] $\lambda < 0$ and $m^2 <0$ case \\ \\
In this parameter region, we have 
\begin{align}
&C_1 = \frac{3\mu-\sqrt{9\mu^2+12|m^2| |\lambda|}}{2|\lambda|}<0\, , \quad C_2=\frac{3\mu+\sqrt{9\mu^2+12|m^2||\lambda|}}{2 |\lambda|}>0 \, , \nn
&C_- = \frac{2 \mu - \sqrt{4\mu^2 + 6 |m^2| |\lambda|}}{|\lambda|} <0\, , \quad C_+=\frac{2 \mu + \sqrt{4\mu^2 + 6 |m^2| |\lambda|}}{|\lambda|} >0 \, .
\end{align}
The condition for the negative and positive energy solutions are given by
\begin{align}
C_{-} < C < C_{+} \quad &\mathrm{for} \quad \mbox{Positive energy} \, , \nn
C<C_{-} \  \mathrm{or} \ C_{+} <C \quad &\mathrm{for} \quad 
\mbox{Negative energy} \, .
\end{align}
{
We repeat the analysis similar to the one presented above. 
Thus we will only give the results in the following.}
\begin{enumerate}
\item $C_{+}$ and $C_{2}$ \\
By taking the difference between $C_+$ and $C_2$, we find
\begin{align}
C_+ -C_2&=\frac{\mu}{2 |\lambda|} \left[ 1 +4 \sqrt{1+\frac{3|m^2||\lambda|}{2\mu^2}}-3\sqrt{1+\frac{4|m^2||\lambda|}{3\mu^2}} \right] \nn
&> \frac{\mu}{2 |\lambda|} \left[ 1+\sqrt{1+\frac{4|m^2||\lambda|}{3\mu^2}} \right]>0 \, .
\end{align}
This meams that $C_2$ is a positive energy solution because $C_2$ takes a positive value.
\item $C_{-}$ and $C_{1}$ \\
By taking the difference between $C_-$ and $C_1$, we find
\begin{align}
C_- -C_1&=\frac{\mu}{2 |\lambda|} \left[ 1 -4 \sqrt{1+\frac{3|m^2||\lambda|}{2\mu^2}}+3\sqrt{1+\frac{4|m^2||\lambda|}{3\mu^2}} \right] \nn
&< \frac{\mu}{2 |\lambda|} \left[ 1-\sqrt{1+\frac{3|m^2||\lambda|}{2\mu^2}} \right]<0 \, .
\end{align}
Therefore, $C_1$ is a also positive energy solution as $C_1 <0$ is obviously smaller than $C_+ >0$.
\end{enumerate}
\item{(d)} $0<\lambda < 3\mu^2 /4|m^2| $ and $m^2<0$\\ \\
$C_1$, $C_2$ and $C_{\pm}$ are given by 
\begin{align}
&C_1 = \frac{-3\mu+\sqrt{9\mu^2-12|m^2| \lambda}}{2\lambda}<0\, , \quad C_2=\frac{-3\mu-\sqrt{9\mu^2-12|m^2|\lambda}}{2 \lambda}<0 \, , \nn
&C_- = \frac{-2 \mu - \sqrt{4\mu^2 - 6 |m^2| \lambda}}{\lambda} <0\, , \quad C_+=\frac{-2 \mu + \sqrt{4\mu^2 - 6 |m^2| \lambda}}{\lambda} <0 \, .
\end{align}
The energy conditions are
\begin{align}
C_{-} < C < C_{+} \quad & \mathrm{for} \quad \mbox{Negative energy} \, , \nn
C<C_{-} \  \mathrm{or} \ C_{+} <C \quad & \mathrm{for} \quad 
\mbox{Positive energy} \, .
\end{align}
Since $C_{\pm}$ are not real in the case of $2\mu^2/3|m^2| < \lambda < 3 \mu^2 /4 |m^2|$, we devide the 
parameter region as the previous case. \\ \\
\textbf{(d-1)} $0< \lambda < 2\mu^2/3|m^2|$
\begin{enumerate}
\item $C_{+}$ and $C_{1}$ \\
\begin{align}
C_+ -C_1&=\frac{\mu}{2 \lambda} \left[ -1 + 4 \sqrt{1-\frac{3|m^2|\lambda}{2\mu^2}}-3\sqrt{1-\frac{4|m^2|\lambda}{3\mu^2}} \right] \nn
&< \frac{\mu}{2 |\lambda|} \left[ -1+\sqrt{1-\frac{3|m^2||\lambda|}{2\mu^2}} \right]<0 \, .
\end{align}
This means that $C_1$ is a positive energy solution.  
\item $C_{+}$ and $C_{2}$ \\
\begin{align}
C_+ -C_2 &=\frac{\mu}{2 \lambda} \left[ -1 +4 \sqrt{1-\frac{3|m^2|\lambda}{2\mu^2}}+3\sqrt{1-\frac{4|m^2|\lambda}{3\mu^2}} \right] \nn
&> \frac{\mu}{2 \lambda} \left[ -1 +4 \sqrt{1-\frac{3|m^2|\lambda}{2\mu^2}}+3 \cdot \frac{1}{3} \right] >0 \, .
\end{align}
This is because $1/3 < \sqrt{1-4|m^2|\lambda/3\mu^2} <1$.
\item $C_{-}$ and $C_{2}$ \\
\begin{align}
C_- -C_2&=\frac{\mu}{2 \lambda} \left[ -1 - 4 \sqrt{1-\frac{3|m^2|\lambda}{2\mu^2}}+3\sqrt{1-\frac{4|m^2|\lambda}{3\mu^2}} \right] \nn
&< \frac{\mu}{2 \lambda} \left[ -1 - 3 \left( \sqrt{1-\frac{3|m^2|\lambda}{2\mu^2}}-\sqrt{1-\frac{4|m^2|\lambda}{3\mu^2}} \right) \right] \, .
\end{align}
As in the case of (\ref{thy}), the maximum value of the second term is given by 
\begin{align}
\left|\sqrt{1-\frac{3|m^2|\lambda}{2\mu^2}}-\sqrt{1-\frac{4|m^2|\lambda}{3\mu^2}} \right| <\frac{1}{3} \, .
\end{align}
Hence, we find $C_- < C_2$. $2$ and $3$ mean $C_2$ is a negative energy solution.
\end{enumerate}
\textbf{(d-2)} $2\mu^2/3|m^2| < \lambda < 3\mu^2/4|m^2| $ \\ \\
$C_{\pm}$ are not real in this parameter region. Thus, both solutions $C_{1,2}$ correspond to the positive energy.
\end{description}

These results are summarized in the table I and II. The former and the latter correspond to the cases of $m^2>0$
and $m^2 <0$, respectively.\\

As we mentioned, the Fierz-Pauli theory is stable on the local maximum. 
Therefore, it is plausible to assume that the theory is stable on the local maximum even 
though the parameters $\mu$ and $\lambda$ take non-vanishing value. 
Under this assumption, we check the stability of the solution $C_{1,2}$. 
For this purpose, we have to obtain the expression of the second derivative of the 
potential:
\be
V''(C) = -12 m^2 + 24 \mu C + 12 \lambda C^4 \, .
\label{2nd pot}
\ee
We find the stability by substituting the solutions into  (\ref{2nd pot}) for each parameter 
region. 
\begin{description}
\item[(a)] $\lambda >0 $ and $m^2 >0$ \\ \\
\hspace*{-4mm} In this case, both solutions corresponds to the negative energy solutions. 
Plugging these solution yields 
\begin{enumerate}
\item $C=C_1$
\be
V''(C=C_1) = \frac{3}{\lambda} \left(6 \mu^2 - 2 \mu \sqrt{9\mu^2+12m^2 \lambda} \right)
+24m^2 >0 \, .
\ee
\item $C=C_2$ \\
\be
V''(C=C_2) = \frac{3}{\lambda} \left( 6 \mu^2 + 2 \mu \sqrt{9\mu^2+12m^2 \lambda} \right)
+24m^2 >0 \, .
\ee
\end{enumerate}
\hspace*{-4mm} (\ref{2nd pot}) is positive in the both cases. 
Therefore, these solutions are unstable.
\item[(b-1)] $-2\mu^2 / 3 m^2 < \lambda < 0$ and $m^2>0$ \\ \\
\hspace*{-4mm} $C_1$ and $C_{2}$ are linked with the positive energy and the negative 
energy solutions, respectively. 
As the above case, \hspace*{-4mm}  we find  (\ref{2nd pot}) for each solution.
\begin{enumerate}
\item $C=C_1$
\be
V''(C=C_1) = \frac{3}{|\lambda|} \left( -6 \mu^2 + 2 \mu \sqrt{9\mu^2 
 -12m^2 |\lambda|} \right)+24m^2 >0 \, .
\ee
\item $C=C_2$ \\
\be
V''(C=C_2) = \frac{3}{|\lambda|} \left(-6 \mu^2 - 2 \mu \sqrt{9\mu^2
 -12m^2 |\lambda|} \right)+24m^2 < 0 \, .
\ee
\end{enumerate}
\hspace*{-4mm} This result means that $C_1$ corresponding to the negative energy 
solution is unstable while $C_2$ corresponding to \hspace*{-1.2mm}the positive energy 
solution is stable.
\item[(b-2)] $-3\mu^2/4m^2 < \lambda < -2 \mu^2 / 3m^2$ and $m^2 >0$\\ \\
\hspace*{-4mm} The negative energy solution is realized for the both solutions $C_{1}$ 
and $C_{2}$.  
\begin{enumerate}
\item $C=C_1$
\be
V''(C=C_1) = \frac{3}{|\lambda|} \left( -6 \mu^2 + 2 \mu \sqrt{9\mu^2
 -12m^2 |\lambda|} \right)+24m^2 >0 \, .
\ee
\item $C=C_2$ \\
\be
V''(C=C_2) = \frac{3}{|\lambda|} \left( -6 \mu^2 - 2 \mu \sqrt{9\mu^2
 -12m^2 |\lambda|} \right) +24m^2 < 0 \, .
\ee
\end{enumerate}
\hspace*{-4mm} Although both solutions lead to the negative energy solution, $C_{1}$ is  
unstable and the other solution is stable. 
\item[(c)] $\lambda < 0$ and $m^2 <0$\\ \\ 
Both solutions correspond to the positive energy. 
\begin{enumerate}
\item $C=C_1$
\be
V''(C=C_1) = \frac{3}{|\lambda|} \left( -6 \mu^2 + 2 \mu \sqrt{9\mu^2
 +12|m^2| |\lambda|} \right)-24|m^2| < 0 \, .
\ee
\item $C=C_2$ \\
\be
V''(C=C_2) = \frac{3}{|\lambda|} \left( -6 \mu^2 - 2 \mu \sqrt{9\mu^2
 +12|m^2| |\lambda|} \right)-24|m^2| < 0 \, .
\ee
\end{enumerate}
\hspace*{-4mm} The both positive energy solutions are stable. 
\item[(d-1)] $0< \lambda < 2\mu^2 / 3 |m^2|$ and $m^2 <0$\\ \\ 
$C_1$ is a positive energy solution and $C_2$ is a negative energy solution.  
\begin{enumerate}
\item $C=C_1$
\be
V''(C=C_1) = \frac{3}{|\lambda|} \left( 6 \mu^2 - 2 \mu \sqrt{9\mu^2
 -12|m^2| \lambda} \right)-24|m^2| < 0 \, .
\ee
\item $C=C_2$ \\
\be
V''(C=C_2) = \frac{3}{|\lambda|} \left( 6 \mu^2 + 2 \mu \sqrt{9\mu^2
 -12|m^2| \lambda} \right) -24|m^2| > 0 \, .
\ee
\end{enumerate}
\hspace*{-4mm} The positive energy solution is stable while the negative energy solution is unstable.
\item[(d-2)] $2\mu^2 / 3 |m^2| < \lambda < 3 \mu^2 /4 |m^2| $ and $m^2 <0$\\ \\ 
In this parameter region, both solutions correspond to the positive energy. The stability analysis
is same as the previous case because the expressions of $C_{1}$and $C_2$ do not change from (d-1). 
Thus, we find that $C_1$ is stable and $C_2$ is unstable.  

\end{description}

}

The above discussion tells that in the solutions $C_{1,2}$ for both of cases (a) and (c), the values of the potential have the same signature but the stability is different from each other.  
Both of solutions are unstable in the case (a) while there exist the stable solutions in the case (c). The cases (b-1) and (d-1) have one stable positive energy solution and one unstable negative energy solution. These results are summarized in the table I.

We also comment on the global structure of the potential and the global stability for the massive spin two field. 
The special feature in the model of massive spin two particle is that the vacuum where the potential is convex upward is stable but the vacuum where the potential is convex downward is unstable. 
In case that both of $C_1$ and $C_2$ correspond to the stable vacua, however, the system also has the ``trivial'' vacuum $C=0$ which realizes the lowest enegy in the system although the massive spin two particle becomes tachyon around the vacuum. We may think that the system could be ultimately unstable by the quantum tunneling from the stable ``false'' vacua to the unstable ``true'' vacuum. 
In case of the scalar field theory, this speculation  could be true. In case of the massive spin two field, however, it is not clear if the system is unstable or not because the potential does not correspond to the propagating modes, which is not the scalar mode but the massive spin two mode. 
If we consider the tunneling for the massive spin two mode by, say, the WKB approximation, we need to consider inhomogeneous and anisotropic intermediate states, which makes the situation very complex. 
Therefore at least at present, we do not know how we should discuss about the global stability and we only concentrate on the arguments about the local stability.

\begin{table}[]
\caption{The relation between $C_{1,2}$, the vacuum energy and the stability of the solutions when $m^2>0$}\label{tab:solution}
  \begin{tabular}{|c|c|c|c|} \hline 
    \backslashbox{Energy}{parameters} & $\lambda>0 $ & 
$-2\mu^2/3m^2 < \lambda < 0$ & $-3\mu^2/4m^2< \lambda <-2\mu^2/3m^2$ \\ \hline
positive energy & no solution & $C_2$ (stable)& no solution \\ \hline
 negative energy & $C_1$ (unstable) and $C_2$ (unstable)& $C_1$ (unstable) & $C_1$ 
(unstable) and $C_2$ (stable) \\ \hline
\end{tabular}
\end{table}

\begin{table}[]
\caption{The relation between $C_{1,2}$, the vacuum energy and the stability of the solutions when $m^2<0$}\label{tab:solution2}
  \begin{tabular}{|c|c|c|c|} \hline 
    \backslashbox{Energy}{parameters} & $\lambda<0 $ & 
$0 < \lambda < 2\mu^2/3|m^2|  $ & $ 2\mu^2/3|m^2| < \lambda < 3\mu^2/4|m^2|$ \\ \hline
positive energy & $C_1$ (stable) and $C_2$ (stable) & $C_1$ (stable)&  $C_1$ 
(stable) and $C_2$ (unstable)\\ \hline
 negative energy & no solution & $C_2$ (unstable) & no solution \\ \hline
\end{tabular}
\end{table}

\section{New model of massive spin-two particle in a curved spacetime}

The naive extension to the theory in a curved spacetime is  given by the minimal-coupling model. 
\begin{align}
\label{hhhhc}
S =& \int d^4 x \sqrt{-g} \left\{ 
-\frac12 \nabla_\mu h_{\nu \rho} \nabla^\mu h^{\nu \rho} 
+\nabla_\mu h_{\nu \rho } \nabla^\rho h^{\nu \mu}
- \nabla^\mu h_{\mu \nu} \nabla^\nu h 
+\frac12 \nabla_\mu h \nabla^\mu h 
+\frac{m^2}{2} g^{\alpha \beta \gamma \delta } h_{\alpha \beta} h_{\gamma \delta}  \right. \nn
& \left. - \frac{\mu}{3!} g^{\mu_{1} \nu_{1} \mu_{2} \nu_{2} \mu_{3} \nu_{3}} 
h_{\mu_{1}\nu_{1}} h_{\mu_{2} \nu_{2}} h_{\mu_{3} \nu_{3}}
 - \frac{\lambda}{4!} g^{\mu_{1} \nu_{1} \mu_{2} \nu_{2} \mu_{3} \nu_{3} \mu_{4} \nu_{4}} 
h_{\mu_{1} \nu_{1}} h_{\mu_{2} \nu_{2}} h_{\mu_{3} \nu_{3}} h_{\mu_{4} \nu_{4}}
 \right\}\, .
\end{align}
Unfortunately, this minimal coupling model is not ghost-free even in case of the free theory according 
to \cite{Buchbinder:1999ar,Buchbinder:1999be}. Therefore, we constructed a new ghost-free massive spin-two model coupled with 
gravity by adding non-minimal coupling terms \cite{Akagi:2014vua}. Instead of (\ref{hhhhc}), the action of the ghost-free model 
in arbitrary dimensions $D$ is given by
\begin{align}
\label{GfreeM}
S=& \int d^D x\sqrt{-g} \left\{ -\frac12 \nabla_\mu h_{\nu \rho} \nabla^\mu h^{\nu \rho} 
+\nabla_\mu h_{\nu \rho } \nabla^\rho h^{\nu \mu}
- \nabla^\mu h_{\mu \nu} \nabla^\nu h 
+\frac12 \nabla_\mu h \nabla^\mu h  +\frac{m^2}{2}g^{\alpha \beta \gamma \delta} h_{\alpha \beta}h_{\gamma \delta} \right. 
\nn
& \left.+\frac{\xi}{D} R h_{\mu\nu} h^{\mu\nu} +\frac{1-2\xi}{2D} R h^2 
- \frac{\mu}{3!} g^{\mu_{1} \nu_{1} \mu_{2} \nu_{2} \mu_{3} \nu_{3}} 
h_{\mu_{1}\nu_{1}} h_{\mu_{2} \nu_{2}} h_{\mu_{3} \nu_{3}} 
- \frac{\lambda}{4!} g^{\mu_{1} \nu_{1} \mu_{2} \nu_{2} \mu_{3} \nu_{3} \mu_{4} \nu_{4}} 
h_{\mu_{1} \nu_{1}} h_{\mu_{2} \nu_{2}} h_{\mu_{3} \nu_{3}} h_{\mu_{4} \nu_{4}}
\right\} \, .
\end{align}
In addition to two non-minimal coupling terms, we also found the following 
non-derivative interaction terms in four dimensions \cite{Akagi:2014vua}.
\begin{align}
\label{Weyl}
& C^{\mu_1 \mu_2 \nu_1 \nu_2} h_{\mu_1 \nu_1} h_{\mu_2 \nu_2 } \, , \nn
&\delta^{\mu_1 ~~\mu_2 ~~ \mu_3}_{~~\rho_1 ~~\rho_2 ~~\rho_3 } 
 \delta^{\nu_1 ~~\nu_2 ~~ \nu_3}_{~~\sigma_1 ~~\sigma_2 ~~\sigma_3 } 
C^{\rho_1 \rho_2 \sigma_1 \sigma_2} g^{\rho_3 \sigma_3} h_{\mu_1 \nu_1} h_{\mu_2 \nu_2} 
h_{\mu_3 \nu_3} \, , \nn
& \delta^{\mu_1 ~~\mu_2 ~~ \mu_3 ~~\mu_4}_{~~\rho_1 ~~\rho_2 ~~\rho_3 ~~\rho_4} 
 \delta^{\nu_1 ~~\nu_2 ~~ \nu_3 ~~\nu_4}_{~~\sigma_1 ~~\sigma_2 ~~\sigma_3 ~~\sigma_4} 
C^{\rho_1 \rho_2 \sigma_1 \sigma_2} g^{\rho_3 \sigma_3 \rho_4 \sigma_4} 
h_{\mu_1 \nu_1} h_{\mu_2 \nu_2}  h_{\mu_3 \nu_3} h_{\mu_4 \nu_4} \, .
\end{align}
Here $C_{\mu\nu\rho\sigma}$ is the Weyl tensor defined by
\be
\label{Weyldef}
C_{\mu\nu\rho\sigma} 
\equiv R_{\mu\nu\rho\sigma} - \frac{1}{2} \left( g_{\mu\rho} R_{\nu\sigma} + g_{\nu\sigma} R_{\mu\rho} 
 - g_{\mu\sigma} R_{\nu\rho} - g_{\nu\rho} R_{\mu\sigma} \right)
+ \frac{1}{6} R \left( g_{\mu\rho} g_{\nu\sigma} - g_{\mu\sigma} g_{\nu\rho} \right)\, .
\ee
Note that the interaction terms containing the scalar curvature like $R^n g^{\mu_1 \nu_1 \mu_2 \nu_2 \mu_3 \nu_3}h_{\mu_1 \nu_1}h_{\mu_2 \nu_2}h_{\mu_3 \nu_3}$
can be added because $R$ is constant on the Einstein manifold, but we ignore such a redundant term here.

\section{Classical solutions and stability}
\label{CSS}

In the previous section, we revealed the parameter region which allows the system to have stable solutions. 
Although the result also important, the analysis is not enough because of the appearance
of the non-minimal coupling term.

In this analysis, we assume the four dimensional (anti)-de Sitter spacetime as a background metric where the non-minimal coupling terms containing 
the Weyl tensor (\ref{Weyl}) vanish. Therefore,  we consider the action (\ref{GfreeM}) in four dimensions.
\begin{align}
\label{GfreeM4}
S=& \int d^4 x\sqrt{-g} \left\{ -\frac12 \nabla_\mu h_{\nu \rho} \nabla^\mu h^{\nu \rho} 
+\nabla_\mu h_{\nu \rho } \nabla^\rho h^{\nu \mu}
- \nabla^\mu h_{\mu \nu} \nabla^\nu h 
+\frac12 \nabla_\mu h \nabla^\mu h  +\frac{m^2}{2}g^{\alpha \beta \gamma \delta} h_{\alpha \beta}h_{\gamma \delta} \right. 
\nn
& \left.+\frac{\xi}{4} R h_{\mu\nu} h^{\mu\nu} +\frac{1-2\xi}{8} R h^2 
- \frac{\mu}{3!} g^{\mu_{1} \nu_{1} \mu_{2} \nu_{2} \mu_{3} \nu_{3}} 
h_{\mu_{1}\nu_{1}} h_{\mu_{2} \nu_{2}} h_{\mu_{3} \nu_{3}} 
- \frac{\lambda}{4!} g^{\mu_{1} \nu_{1} \mu_{2} \nu_{2} \mu_{3} \nu_{3} \mu_{4} \nu_{4}} 
h_{\mu_{1} \nu_{1}} h_{\mu_{2} \nu_{2}} h_{\mu_{3} \nu_{3}} h_{\mu_{4} \nu_{4}}
\right\} \, .
\end{align}
 
As in the case of the flat spacetime, vacuum solutions have to be invariant under the isometry of the spacetime. 
\be
\label{ansatzC}
h_{\mu \nu} = C g_{\mu \nu}\, , \qquad \mathcal{L}_{\xi} g_{\mu \nu} =0 \, . 
\ee
Here $C$ is a constant and $\xi$ is the Killing vectors for the (anti-) de Sitter spacetime. 
Substituting the ansatz (\ref{ansatzC}) into the equations of motion gives
\be
 - 2 \left\{ 6m^2 + \left( 2-3\xi \right)R \right\} C + 12 \mu C^2 + 4 \lambda C^3 =0 \, .
\ee
This is the equation determining the extrema of the potential for the system. The solutions are 
given as follows:
\begin{align}
C=0\, , \qquad C=\frac{-6 \mu \pm \sqrt{36 \mu^2 +48 m^2 \lambda + 8 \lambda (2-3\xi) R }}{4 \lambda} \, .
\end{align}
The condition for the existence of the non-trivial solutions is 
\be
\label{CES}
9\mu^2+12m^2 \lambda + 2 \lambda (2-3\xi)R>0\, .
\ee
In order to investigate the stability around the vacuum solution, we consider the fluctuation 
\be
h_{\mu \nu} = C g_{\mu \nu} +f_{\mu \nu} \, ,
\ee
and rewrite the action in terms of $f_{\mu \nu}$.
\begin{align}
\label{Ak37}
S = & \int d^4x \sqrt{-g} 
\left\{ -\frac12 \nabla_\mu f_{\nu \rho} \nabla^\mu f^{\nu \rho}
+\nabla_\mu f_{\nu \rho } \nabla^\rho f^{\nu \mu}
- \nabla^\mu f_{\mu \nu} \nabla^\nu f 
+\frac12 \nabla_\mu f \nabla^\mu f 
 -V(C) \right\}
\, .
\end{align}
Here $V(C)$ takes the following form:
\begin{align}
\label{Ak38}
V(C) = & \sum_{n=0}^4V_n(C) \, , \nn
V_0(C) = & - \left\{ 6m^2 + \left( 2-3\xi \right)R \right\} C^2 
+ 4\mu C^3 + \lambda C^4 \, , \nn
V_1(C) = & -3m^2 Cf+ 3\mu C^2 f +\lambda C^3 f - \left( 1-\frac{3}{2} \xi \right) RCf 
=\frac{{V'}_0(C)}{4} f \, , \nn
V_2(C) =& \left(-\frac{m^2}{2} +\mu C + \frac{\lambda}{2} C^2 \right) 
g^{\alpha \beta \gamma \delta}f_{\alpha \beta} f_{\gamma \delta}
 -\frac{\xi}{4} R f_{\alpha \beta} f^{\alpha \beta} -\frac{1-2\xi}{8} R f^2 \, .  \\
& \vdots \nonumber
\end{align}
Here $V_n(C)$ expresses the term including $n$-the power of $f_{\mu\nu}$. 
We should note that $V_2(C)$ is proportional to the Fierz-Pauli mass term 
in (\ref{FPLag}) due to the following identity,
\be
\label{Ak39}
{g^{\mu_1 \nu_1 \cdots \mu_{n-1} \nu_{n-1} \mu_n}}_{\mu_n} 
= (D-n+1) g^{\mu_1 \nu_1 \cdots \mu_{n-1} \nu_{n-1}}\, .
\ee
Here $D$ denotes the dimensions of the spacetime. 
We should also note that $V_3(C)$ and $V_4(C)$ are also given by the pseudo linear 
terms in (\ref{nhhh}) and (\ref{nhhhh}). 

We now define an effective mass $M$ of $f_{\mu\nu}$ by 
$M^2 \equiv m^2 -2\mu C -\lambda C^2 $. As the vacuum solutions satisfy the equation ${V'}_0 (C)=0$, the linear term in $f_{\mu \nu}$ vanishes.
\begin{align}
S =& \int d^4x \sqrt{-g} \left\{ - \frac{1}{2} \nabla_\mu f_{\nu \rho} \nabla^\mu f^{\nu \rho} 
+\nabla_\mu f_{\nu \rho } \nabla^\rho f^{\nu \mu}
- \nabla^\mu f_{\mu \nu} \nabla^\nu f 
+\frac{1}{2} \nabla_\mu f \nabla^\mu f 
 \right. \nn
& \left. +\frac{M^2}{2} g^{\alpha \beta \gamma \delta} f_{\alpha \beta} f_{\gamma \delta} 
+\frac{\xi}{4} R f_{\alpha \beta} f^{\alpha \beta} +\frac{1-2\xi}{8} R f^2
 - V_0 (C) +  \mathcal{O} \left(f^3, f^4\right) \right\} \, .
\label{eq7}
\end{align}
Because the purpose is to investigate the stability around the vacua, we need to keep the terms including the second power of $f_{\mu\nu}$. 
In a curved spacetime, the stability of the free massive spin-two field is determined by the Higuchi bound \cite{Higuchi:1986py}. 
In case that $\lambda=\mu=0$ and $\xi=1$, it is well known that by assuming  $M^2 = m^2 >0$, if $R>6m^2$, the vacuum is unstable 
and if $R\le 6m^2$, stable \cite{Higuchi:1986py,Deser:2001wx}. 
On the boundary $R=6 m^2$, the theory is invariant under the gauge transformation 
\begin{equation}
\delta f_{\mu \nu}=\nabla_{\mu} \nabla_{\nu} \Gamma +\frac{1}{2} M^2 \Gamma \nonumber
\, ,
\end{equation}
where $\Gamma$ is a gauge parameter. For this reason, the theory satisfying the condition $R=6m^2$ with $\lambda=\mu=0$ and $\xi=1$ is called partially massless.
The stability has not been investigated when $\xi \neq 1$ but {
as we see below, the deviation is not very important when the curvature of the spacetime is constant. 
Let us see the quadratic term in the potential of (\ref{GfreeM4}):
\begin{align}
\label{Rmass}
-\frac{m^2}{2} (h^{\mu \nu} h_{\mu \nu}-h^2) +\frac{\xi}{4} R h_{\alpha \beta} h^{\alpha \beta} +\frac{1-2\xi}{8} R h^2\, .
\end{align}
To address the deviation from the $\xi=1$ case, we express the $\xi$ parameter in terms of $\delta$:
\be
\label{pd}
\xi=1+\delta \, .
\ee
Then, we rewrite (\ref{Rmass}) as follows:
\begin{align}
\label{reda}
-\frac{m^2}{2}(h^{\mu \nu}h_{\mu \nu}-h^2) + \frac{1}{4}Rh^{\mu \nu}h_{\mu \nu}-\frac{1}{8}Rh^2+\frac{\delta}{4}R(h^{\mu \nu}h_{\mu \nu}-h^2) \, .
\end{align}
This means that the deviation from $\xi=1$ is equivalent to the shift in the mass parameter since $R$ is constant and we can set $\xi=1$ without loss of generality.
(We should note that this is just a mathematical equivalence. Because the mass parameter is strongly related with the stability of the system, 
the deviation from $\xi=1$ is physically important.) Hence, we {
impose} the following conditions for the stability,
\begin{align}
\label{Ak40}
M^2 \geq 0 \, , \quad 
R \le 6M^2 \, .
\end{align}
Therefore, the stable, nontrivial solutions have to satisfy both of the conditions (\ref{CES}) and (\ref{Ak40}). 
For example, let us consider the simple case where $\mu=0$. The non-trivial solution is simplified as follows:
\be
\label{SimV}
C= \pm \sqrt{\frac{6m^2 -R}{2 \lambda}} \, .
\ee
The condition for the existence of (\ref{SimV}) is  $\lambda < 0$ and $ R>6m^2$ or $\lambda > 0$ and $ R<6m^2$.
On the other hand, the effective mass $M$ around the vacuum is given by
\be
\label{EfM}
M=m^2 - \lambda C^2 = -2m^2+\frac{R}{2} \, .
\ee
 From (\ref{Ak40}) and (\ref{EfM}), we have the stability condition around the vacuum expectation value (VEV) as follows, 
\be
\label{HV}
R > 4m^2, \quad R> 6m^2 \, .
\ee
The stable region is shown in FIG 1. Hence, the solution satisfying the stability condition exists when $\lambda < 0$ and $ R>6m^2$.
\begin{figure}[t]
\begin{center}
\includegraphics[width=100mm]{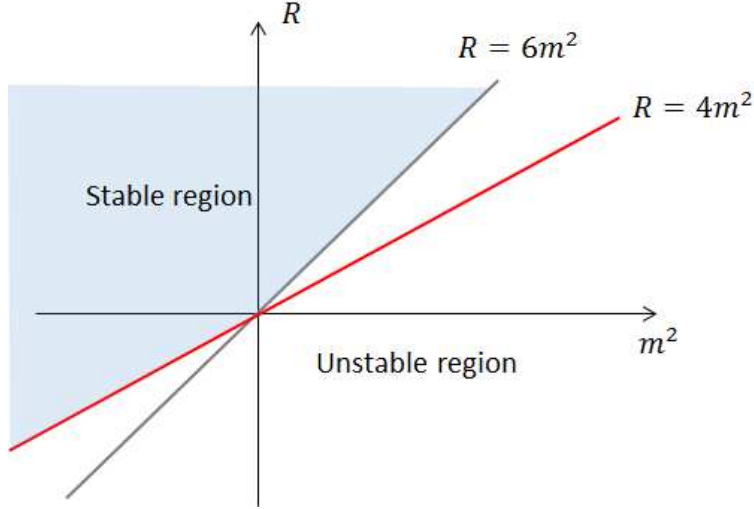}
\end{center}
\caption{The parameter region for the constant curvature spacetime.}
\end{figure}

}


\section{New bigravity}

The bigravity model can be regarded as a model where massive spin-two field couples with 
gravity. Then we may consider the model where $h_{\mu\nu}$ couples with gravity which can 
be regarded as a new bigravity model because there appear two symmetric 
tensor fields $g_{\mu\nu}$ and $h_{\mu\nu}$, as follows, 
\begin{align}
\label{Ak27}
S=& \int d^4 x\sqrt{-g} \left\{ -\frac12 \nabla_\mu h_{\nu \rho} \nabla^\mu h^{\nu \rho} 
+\nabla_\mu h_{\nu \rho } \nabla^\rho h^{\nu \mu}
- \nabla^\mu h_{\mu \nu} \nabla^\nu h 
+\frac12 \nabla_\mu h \nabla^\mu h  +\frac{m^2}{2}g^{\alpha \beta \gamma \delta} h_{\alpha \beta}h_{\gamma \delta} \right. 
\nn
& \left.+\frac{\xi}{4} R h_{\mu\nu} h^{\mu\nu} +\frac{1-2\xi}{8} R h^2 
- \frac{\mu}{3!} g^{\mu_{1} \nu_{1} \mu_{2} \nu_{2} \mu_{3} \nu_{3}} 
h_{\mu_{1}\nu_{1}} h_{\mu_{2} \nu_{2}} h_{\mu_{3} \nu_{3}} 
- \frac{\lambda}{4!} g^{\mu_{1} \nu_{1} \mu_{2} \nu_{2} \mu_{3} \nu_{3} \mu_{4} \nu_{4}} 
h_{\mu_{1} \nu_{1}} h_{\mu_{2} \nu_{2}} h_{\mu_{3} \nu_{3}} h_{\mu_{4} \nu_{4}}
\right\} + S_\mathrm{EH} \, .
\end{align}
Here $h_{\mu\nu}$ is not the perturbation in $g_{\mu\nu}$ but 
$h_{\mu\nu}$ is a field independent of $g_{\mu\nu}$ and $S_\mathrm{EH}$ is given by the Einstein-Hilbert action
without the cosmological constant. We have to note that the action is constructed up to the dimension-4 operators and it is not obvious whether or
not this system has a ghost
since the matter part of the Lagrangian is constructed for the field $h_{\mu \nu}$ to be ghost-free only on the Einstein manifold.
We show, however, that there are classical solutions which realize the spacetime which has constant curvature.

We also stress that the $\xi$ parameter is not redundant here unlike the rigid curved background. 
In the section \ref{CSS}, we see 
that the deviation from $\xi=1$ means the appearance of the Fierz-Pauli tuned term  (\ref{reda}) proportional to the constant curvature $R$. 
Thus, we concluded such a term is redundant and can be ignored without loss of generality. On the other hand, in (\ref{Ak27}), as $R$ is not constant 
but a dynamical variable, we cannot regard the $\xi$ as a redundant parameter. 

Finally, we note that the mass parameter $m$, the cubic coupling $\mu$ and the quartic coupling $\lambda$ take an arbitrary real value.

\section{Cosmological and Black hole Solutions}


In this section, we obtain the cosmological solution and the black hole solutions. We now assume, for simplicity,that the solution 
has the following form as in the section \ref{CSS},  
\begin{align}
\label{Ak28}
h_{\mu \nu} = C g_{\mu \nu}\, .
\end{align}
Here $C$ is a constant. The equations of motion for $h_{\mu \nu}$ are given by
\be
\label{CEMH}
 - 2 \left\{ 6m^2 + \left( 2-3\xi \right)R \right\} C + 12 \mu C^2 + 4 \lambda C^3 =0 \, .
\ee
Furthermore, we obtain the following action by substituting (\ref{Ak28}) into (\ref{Ak27}).
\begin{align}
S =& -\int d^4x \sqrt{-g} 
 V_0 (C)
+\frac{1}{2\kappa^2} \int d^4x\sqrt{-g} R \nn
=& \left\{ \left( 2-3\xi \right) C^2+\frac{1}{2\kappa^2} \right\} 
\int d^4 x \sqrt{-g} \left[ R -2 \Lambda_\mathrm{eff} \right] \, .
\label{eqq7}
\end{align}
Here we assume $\left( 2-3\xi \right) C^2+\frac{1}{2\kappa^2} \neq 0$; otherwise
the gravity completely decouples. The effective cosmological constant $\Lambda_\mathrm{eff}$ is defined by
\be
\Lambda_\mathrm{eff} \equiv \frac{\kappa^2 \left(-6m^2 C^2 + 4\mu C^3 
+ \lambda C^4 \right)}{2\kappa^2 C^2 \left(2-3 \xi \right) +1 } \, .
\label{eq5}
\ee 
Thus, the Einstein equation given by the variation of the 
action with respect to $g_{\mu \nu}$ has the following form: 
\begin{align}
R_{\mu \nu} - \frac{1}{2} R g_{\mu \nu} +\Lambda_\mathrm{eff} g_{\mu \nu} = 0 \, .
\label{eq1}
\end{align}
Since (\ref{eq1}) admits the Einstein manifolds as solutions, we can consider the fluctuation around the solution of (\ref{CEMH}) without any ghost
as far as the fluctuation is small enough.
By multiplying (\ref{eq1}) with $g^{\mu\nu}$, we obtain
\begin{align}
R = 4 \Lambda_\mathrm{eff} \, .
\label{eq3}
\end{align}
By substituting (\ref{eq3}) into the expression of (\ref{CEMH}), we find
\be
\label{Ak42}
4C \left\{ -2\mu \zeta C^3 + \left(\lambda + 6\zeta m^2\right) C^2 
+3\mu C -3m^2 \right\} =0 \, .
\ee
Here $\zeta \equiv \kappa^2 \left( 2-3\xi \right)$.
Then the solutions except the trivial solution $C=0$ should satisfy the following condition, 
\be
 -2\mu \zeta C^3 + \left( \lambda + 6\zeta m^2 \right) C^2 +3\mu C -3m^2 =0 \, .
\label{eq4}
\ee
Dividing (\ref{eq4}) by $-2\mu \zeta$ and changing the variable $C$ by 
$C=x+\frac{\lambda+ 6\zeta m^2 }{6\mu \zeta }$, we can rewrite (\ref{eq4}) 
as follows, 
\be
\label{Ak44}
x^3 + px +q =0 \, , \quad 
p= -\frac13 \left\{ \left( \frac{\lambda + 6\zeta m^2}{2\mu \zeta} \right)^2 
+\frac{9}{2\zeta} \right\} \, , \quad 
q= \frac{2}{27} \left( \frac{\lambda + 6\zeta m^2}{2\mu \zeta} \right)^3 
 -\frac{\lambda}{4\mu \zeta^2} \, .
\ee
Then by putting $\omega \equiv \e^{i 2\pi /3}$, the solutions of (\ref{Ak44}) are 
expressed as 
\be
\label{Ak45}
x= \omega^k \sqrt[3]{-\frac{q}{2} + \sqrt{\left( \frac{q}{2} \right)^2 
+ \left( \frac{p}{3} \right)^3} } 
+ \omega^{3-k} \sqrt[3]{-\frac{q}{2} - \sqrt{\left( \frac{q}{2} \right)^2 
+ \left( \frac{p}{3} \right)^3} } \, , \quad k=1,2,3\, .
\ee
Now the determinant is given by 
\be
\label{Ak46}
D= -27q^2 -4 p^3 =-2^2 \cdot 3^3 \left\{ \left( \frac{q}{2} \right)^2 
+ \left( \frac{p}{3} \right)^3 \right\} \, .
\ee
Except the case that $q=p=0$, there are following three cases: 
\begin{enumerate}
\item\label{case1} $D>0$ There are three different real solutions. 
\item\label{case2} $D<0$ There is only one real solution. 
\item\label{case3} $D=0$ There are three real solutions but two of them are degenerate 
with each other. 
\end{enumerate}
Let us consider a little bit simple case $\mu=0$ in the following. 
The equations of motion (\ref{eq4}) is reduced to be
\begin{align}
\label{Ak47}
\left( \lambda +6\zeta m^2 \right) C^2 -3m^2 =0 \, .
\end{align}
Then the solutions are given by, 
\begin{align}
\label{Ak48}
C_1 = \sqrt{\frac{3m^2}{\lambda + 6\zeta m^2}}\, , \quad 
C_2 = -\sqrt{\frac{3m^2}{\lambda + 6\zeta m^2}} \, .
\end{align}
The solutions become real and non-trivial ($C\neq 0$) when 
\be
\label{eq6}
\begin{cases}
m^2>0\, , \quad \lambda+6\zeta m^2>0 \\
m^2<0\, , \quad \lambda+6\zeta m^2<0  
\end{cases} \, .
\ee
Furthermore by substituting $C_{1,2}$ into the effective cosmological constant (\ref{eq5}), 
we obtain 
\be
\label{Ak50}
\Lambda_\mathrm{eff} (C_{1,2}) = -\kappa^2 (C_{1,2})^2 m^2 \, .
\ee
If the conditions in Eq.~(\ref{eq6}) is satisfied, $C_{1,2}$ are real numbers. 
Therefore we find 
\be
\label{Ak51}
\begin{cases}
m^2>0\, , \quad \lambda+6\zeta m^2>0 & \text{Anti-de Sitter} \\
m^2<0\, , \quad \lambda+6\zeta m^2<0 & \text{de Sitter}
\end{cases} \, .
\ee

As we have obtained the solutions of this new bigravity theory, we have to investigate the stability of the obtained solutions as in the case of the rigid background. For this purpose, we need to consider both of the fluctuations in $h_{\mu \nu}$ and $g_{\mu \nu}$, simultaneously but this leads to very complicated equations. Thus, we will analyze the stability in the future work. 

\section{Black hole entropy}

Since the action (\ref{Ak27}) admits the Schwarzschild-anti de Sitter black hole solution under the assumption $h_{\mu \nu}=C g_{\mu \nu}$,
we can calculate the black hole entropy. Let me use the Wald formula to calculate the entropy for the system, which 
is also applicable for the spacetime having the asymptotically anti de Sitter spacetime. 

(We could also have the Schwarzschild-de Sitter solution, but there are several subtleties due to the presence of the two event horizons.)
Note that the Wald formula is applicable for the asymptotically anti-de Sitter spacetime because the mass can be defined based on the asymptotic
killing vector.

The Wald formula is given by
\be
\label{Wald}
S=-2\pi \oint_{\mathcal{H}} dA \frac{\delta \mathcal{L}}{\delta R_{\mu \nu \rho \sigma}} \epsilon_{\mu \nu} \epsilon_{\rho \sigma} \, .
\ee
Here $\epsilon^{\mu \nu}$ is a binormal tensor and $\mathcal{H}$ denotes the horizon of the black hole. 
The term contributing to the functional derivative is 
\be
{\mathcal{L}}_{\mbox{rel}} = \frac{1}{2 \kappa^2} R + \frac{\xi}{4}R h_{\mu \nu}h^{\mu \nu} +\frac{1-2 \xi}{8} R h^2 \, .
\ee
Therefore, we obtain
\be
\label{fdR}
\frac{\delta \mathcal{L}}{\delta R_{\mu \nu \rho \sigma}}= \left(\frac{\xi}{4} h_{\mu \nu}h^{\mu \nu}+\frac{1-2\xi}{8}h^2 + \frac{1}{2 \kappa^2} \right)\cdot
\frac{1}{2} \left( g^{\mu \rho} g^{\nu \sigma} - g^{\mu \sigma} g^{\mu \rho} \right) \, .
\ee
The substitution of the classical solution $C_{1,2}$ and the Schwarzschild-anti-de Sitter or Kerr-anti-de Sitter metric yields 
\be
S= \frac{A}{4G} + \frac{12 \pi A(2-3\xi) m^2 }{\lambda + 48 \pi G (2-3\xi)m^2}\, .
\ee
The last term corresponds to the contribution from the condensation of the massive spin-two particle.
The area of the event horizon for the Schwarzschild type metric is given by
\begin{align}
A=\frac{16 \pi}{ |\Lambda_{\mathrm{eff}}|} \sinh^2 \left[\frac{1}{3} \sinh^{-1} \left(3 M \sqrt{|\Lambda_{\mathrm{eff}}|} \right) \right]
\, .
\end{align}
Here $M$ denotes the black hole mass.

These results can be compared with those 
\cite{Katsuragawa:2013bma,Katsuragawa:2013lfa} in the Hassan-Rosen bigravity 
model \cite{Hassan:2011zd,Hassan:2011vm,Hassan:2011tf}, where the entropy is given by 
the sum of the contributions from two metric sectors.


\section{Summary}

In this paper, we have investigated the classical solutions for the theories of 
massive spin-two particle in flat spacetime and curved spacetime, which were proposed in 
\cite{Ohara:2014vua, Akagi:2014vua} by coupling the model with gravity.   

In conflict with the intuition, the massive spin-two particle becomes tachyon on the local 
minimum of the potential and the particle is stable on the local maximum, that is, the local 
minimum induces the instability although the local maximum corresponding to the stability. Based on
this analysis, we classified the stable or unstable parameter region for the massive spin-two particle 
with potential terms in a flat spacetime. Although the model is very similar to a scalar field 
theory with quartic and quadratic potential terms, it is remarkable that the relation between the stability 
and the vacuum energy is opposite to the model of the scalar field having the similar potential. 

We extend the stability analysis to the case of the rigid background. In this case, the vacuum 
solutions are invariant under the transformation induced by the Killing vector for (anti-) de Sitter spacetime.
Since the stability condition called the Higuchi bound for the free massive spin-two particle is given by Higuchi \cite{Higuchi:1986py}, we apply 
the analysis to our model.    

Finally, we consider the case where the background metric is dynamical due to the presence of the Einstein-Hilbert term.
Then we obtained solutions describing the (anti)-de Sitter spacetime. The obtained de Sitter spacetime might correspond to
the inflation in the early universe or the accelerating expansion in the present universe. 
These solutions correspond to the extrema of the potential for the trace of the symmetric tensor field. As mentioned 
in the text, we do not carry out the stability analysis for this gravity coupled system. This could be a future work. 

In addition to the solutions describing the (anti)-de Sitter spacetime, we find the 
solutions describing the black hole, which could be the (anti)-de Sitter-Schwarzschild or 
the (anti)-de Sitter-Kerr spacetime. By calculating the black hole entropy, furthermore, we find that the entropy contains the explicit 
contribution from the condensation of the massive spin-two particle. 
In case of the Hassan-Rosen bigravity model \cite{Hassan:2011zd,Hassan:2011vm,Hassan:2011tf}, the entropy 
is given by the sum of the contributions from two metric sectors. On the other hand, the black hole entropy for the model in this paper
is not unique because of arbitrary parameters appearing in the entropy.


\section*{Acknowledgments} 

We are indebted to T.~Katsuragawa for the useful discussions. 
The work is supported by the JSPS Grant-in-Aid for Scientific 
Research (S) \# 22224003 and (C) \# 23540296 (S.N.).

\end{document}